\documentclass[twocolumn, nofootinbib, floatfix]{revtex4-1}

\usepackage{amsmath}
\usepackage{graphicx}
\usepackage{dcolumn}
\usepackage{bm}
\usepackage{epsfig}
\usepackage{amssymb,latexsym,mathrsfs}
\usepackage{graphicx}
\usepackage{color}
\usepackage{hyperref}

\hypersetup{
    colorlinks=true,
    linkcolor=red,
    citecolor=blue,
} 

\newcommand{\be}{\begin{equation}}
\newcommand{\ee}{\end{equation}}
\newcommand{\bs}{\begin{split}} 
\newcommand{\bea}{\begin{eqnarray}}
\newcommand{\eea}{\end{eqnarray}}

\newcommand{\ode}{\Omega_{de}}

\newcommand{\geff}{G_{\rm eff}}
\newcommand{\lcdm}{$\Lambda$CDM} 
\newcommand{\rhoi}{\rho_{\pi,i}}
\newcommand{\odet}{\Omega_{de,0}}

\begin{document}

\title{Trial of Galileon gravity by cosmological expansion and growth 
observations} 
\author{Stephen Appleby$^{1}$ \& Eric V.\ Linder$^{1,2}$} 
\affiliation{$^1$Institute for the Early Universe WCU, Ewha Womans University, 
Seoul, Korea\\ 
$^2$Berkeley Center for Cosmological Physics \& Berkeley Lab, 
University of California, Berkeley, CA 94720, USA}

\begin{abstract}
Galileon gravity is a robust theoretical alternative to general relativity 
with a cosmological constant for explaining cosmic acceleration, with 
interesting properties such as having second order field equations and 
a shift symmetry.  While either its predictions for the cosmic expansion or 
growth histories can approach standard $\Lambda$CDM, we demonstrate the 
incompatibility of both doing so simultaneously.  Already current 
observational constraints can severely disfavor an entire class of 
Galileon gravity models that do not couple directly to matter, ruling them out as an alternative to 
$\Lambda$CDM. 
\end{abstract}

\date{\today} 

\maketitle

\section{Introduction} 

General relativity is an excellent description of gravitation on all 
scales at which it has been tested, from the solar system to compact objects 
to cosmology.  However within cosmology, general relativity requires 
a cosmological constant or some form of strongly negative pressure 
to explain observations of late time cosmic acceleration 
\cite{perlmutter98,riess99,komatsu11,araafrieman,araacaldwell}.  
No compelling explanation exists for the magnitude of the cosmological 
constant (or scalar field potential), and in general such 
a contribution to the gravitational action should receive 
corrections from high energy physics. 

These fine tuning and naturalness issues motivate exploration of 
further physics that can explain the observed acceleration and cosmic 
gravity, while being protected against high energy radiative corrections. 
One of the most successful such theories is Galileon gravity 
\cite{nicolis,deffayet1,deffayet2}.  In the standard, cosmological 
constant free case this involves four terms in a Lagrangian that leads to 
well behaved, second order equations of motion.  The Galileon field 
arises as a geometric object in higher dimensions and acts in 4D like a 
shift symmetric scalar field (dark energy), protected against radiative 
corrections in the absence of matter. 
When introducing matter, a coupling is generically expected to arise which 
would break the shift symmetry and spoil the radiative stability of the model. 
However throughout this work we concentrate 
solely on a Galileon scalar field that does not couple explicitly to matter (we call such a model
the `uncoupled' Galileon).

In this {\it Letter\/} we confront this robust theoretical alternative 
to general relativity with current observational data.  This serves as a 
significant example of advancing application of cosmological data to 
probe gravity.  One of the leading alternatives, Galileon gravity 
specifically has intriguing cosmological 
properties such as evolution from a high redshift matter dominated attractor 
to current acceleration and a future de Sitter attractor, and having a 
time varying effective Newton's constant $\geff$.  These cosmological 
characteristics were investigated in detail in \cite{applin} (the existence 
of an asymptotic de Sitter state was first found in \cite{DeFelice:2010pv}); 
numerous other studies \cite{defeliceLCDM,Ali:2010gr,DeFelice:2011aa,Babichev:2011iz,DeFelice:2010as,Mota:2010bs,Brax:2011sv,iorio} 
have examined various properties of Galileon gravity.  Despite its 
attractive theoretical properties, though, does Galileon gravity remain 
viable observationally -- what is the probative power of current data? 

We begin by an analytic discussion of the general dependencies of the 
effective dark energy equation of state and gravitational strength on 
the Galileon parameters.  These will exhibit a tension between the 
cosmic expansion and growth trends, so we proceed numerically with a full 
Markov Chain Monte Carlo scan through the Galileon parameter space, 
comparing the theoretical predictions to the latest cosmological data sets.

\section{Cosmological Properties of Galileons} \label{sec:depend} 

The Galileon action is that of a scalar field $\pi$ non-linearly and derivatively coupled
to itself, and in curved spacetime to the Ricci tensor and its contractions. The action leads to 
field equations no higher than second order (and is hence a subset of Horndeski's theory \cite{WH}), and is invariant under the symmetry
$\pi \to \pi + c + b_{\mu}x^{\mu}$ in the flat space limit, for constant $c$ and $b_{\mu}$.
These conditions results in four invariant combinations \cite{deffayet1} 
\begin{eqnarray} 
& & {\cal L}_{2} = (\nabla_{\mu}\pi)(\nabla^{\mu}\pi), \qquad 
{\cal L}_{3} = (\Box \pi) (\nabla_{\mu}\pi)(\nabla^{\mu}\pi)/M^{3} \\ 
& & {\cal L}_{4} = (\nabla_{\mu}\pi)(\nabla^{\mu}\pi)\left[ 2(\Box\pi)^{2} - 
2 \pi_{;\mu\nu}\pi^{;\mu\nu} - R (\nabla_{\mu}\pi)(\nabla^{\mu}\pi)/2 \right]/M^{6} \\ 
& & {\cal L}_{5} = (\nabla_{\mu}\pi)(\nabla^{\mu}\pi)\left[ (\Box\pi)^{3} - 3(\Box \pi)\pi_{;\mu\nu}\pi^{;\mu\nu} + 2 \pi_{;\mu}{}^{;\nu}\pi_{;\nu}{}^{;\rho}\pi_{;\rho}{}^{;\mu} \right. \\ 
& &\qquad \left. - 6\pi_{;\mu}\pi^{;\mu\nu}\pi^{;\rho}G_{\nu\rho} \right] /M^{9} 
\end{eqnarray} 
where $R$ is the Ricci scalar, $G_{\nu\rho}$ the Einstein tensor, and 
$M^{3} = M_{\rm pl}H_{0}^{2}$ with $M_{\rm pl}$ the Planck mass and $H_0$ 
the Hubble constant.  The full action is then 
\begin{equation}\label{eq:m1}  
S = \int d^{4}x\,\sqrt{-g} \left[ \frac{M_{\rm pl}^{2} R}{2} - 
\frac{1}{2}\sum_{i=2}^{5} c_{i}{\cal L}_{i}   - {\cal L}_{\rm m} \right]  
\end{equation} 
where $c_{2-5}$ are arbitrary dimensionless constants, $g$ is the 
determinant of the metric, and ${\cal L}_{\rm m}$ is the matter Lagrangian that contains no
$\pi$ dependence in this work. 
Generalization of the coefficients to be functions of the field and its 
canonical kinetic term is possible \cite{gengal1,gengal2,horndeski}, but we 
consider the standard Galileon case where the coefficients are constants. 

First, let us explore the broad effects of the Galileon parameters. 
At high redshift, as discussed by \cite{applin}, the effective dark energy 
equation of state $w(z)$ follows 
tracker trajectories given by the background equation of state, i.e.\ 
radiation or matter domination, and so is independent of the parameters. 
The strength of the gravitational coupling $\geff$, however, deviates 
from Newton's constant $G_N$ by an amount proportional to the dark energy 
density at the time, $\ode$.  Thus a key early parameter is the initial 
dark energy density $\rho_{\pi,i}(c_n,H_i,x_i)$, where 
$x=d(\pi/M_{pl})/d\ln (1+z)$ is the field velocity and $z$ the cosmic 
redshift. 

Analytically, increasing $\rho_{\pi,i}$ increases the gravitational strength. 
A substantial increase in the gravitational strength would enhance the 
growth of structure, even at later times, enough to make it discrepant 
with observations.  So we expect that 
growth constraints would favor low values of $\rhoi$, keeping 
$\geff\approx G_N$ for the matter dominated era. 

[More technically, since $\rho_{\pi,i}$ is a function of $c_n$, this 
favors a certain region of the Galileon parameter space.  At high redshift 
the $c_5$ term typically dominates over the others, by factors of 
$H^2 x\gg1$.  For increasingly larger initial densities (and larger $c_5$), 
it takes longer for the other $c_n$ terms to give comparable contributions.  
Since the moderate redshift ($z\approx10$) peak in the gravitational strength 
$\geff$ noted in \cite{applin} occurs 
due to interplay and partial cancellation between the terms, then higher 
values of $\rhoi$ shift the peak to later 
times.  Once the cancellation passes, the peak in the gravitational 
strength often gives way to a period around $z\approx3$ 
where $\geff\approx G_N$ is restored.  Finally, the growth of the 
dark energy density fraction $\ode$ moves $\geff$ instead toward its
late time de Sitter attractor behavior, which is independent of $\rhoi$.  
The basic point, however, is that increasing $\rhoi$ tends to amplify the 
deviation from Einstein gravity, particularly at $z\approx 3-10$.] 

The opposite dependence is true, however, for the expansion constraints.  
If we start with a low $\rhoi$, then due to the approximate tracking 
behavior of $\rho_{\pi}$ during matter domination {\it and\/} the fact 
that we still need to arrive today at $\odet\approx0.7$, one requires a 
more extreme evolution in $w(z)$ near the present.  Analytically, to catch 
up the dark energy density to the present value one must have highly negative 
values of $w(z)$ at $0 \lesssim z \lesssim 2$.  Thus, low $\rhoi$ leads to 
strong spikes in $w(z)$.  
This shifts the distance-redshift relation from the observed, 
near-$\Lambda$CDM behavior, both at low redshift and for the integrated 
distance to the CMB last scattering surface.  

Thus one has a simple analytic picture: Galileon gravity is caught between 
the Scylla of high initial density (and the related region of $c_n$ parameter 
space) pulling $\geff$ and growth unviably up, and the Charybdis of low 
initial density pulling $w(z)$ and distances unviably down.  
(In Homer's {\it Odyssey\/}, Scylla was a cliff-dwelling monster pulling 
sailors up from ships and Charybdis a sea monster sucking them down.)  

\begin{figure}[htbp!]
\includegraphics[angle=-90,width=\columnwidth]{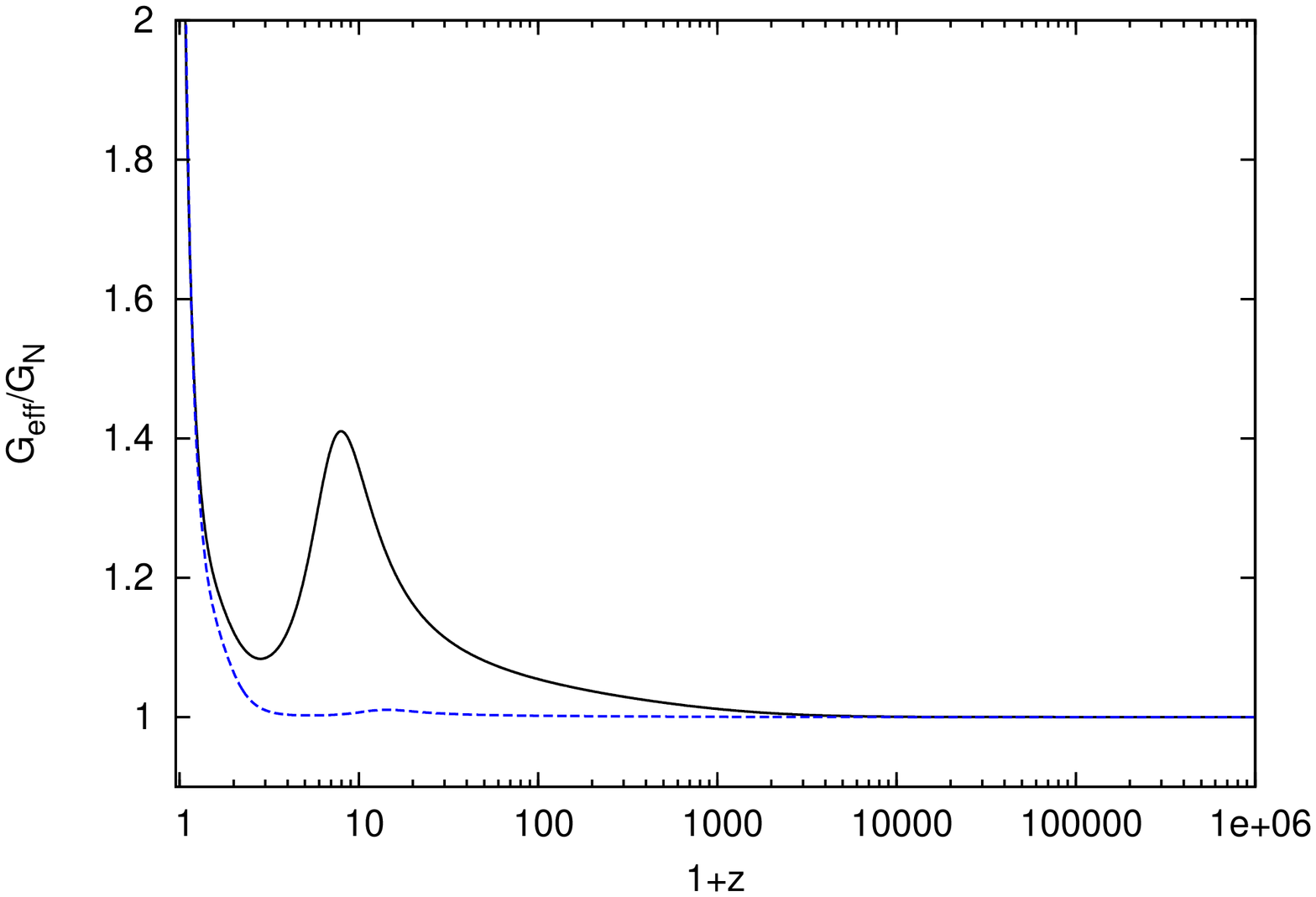}
\includegraphics[angle=-90,width=\columnwidth]{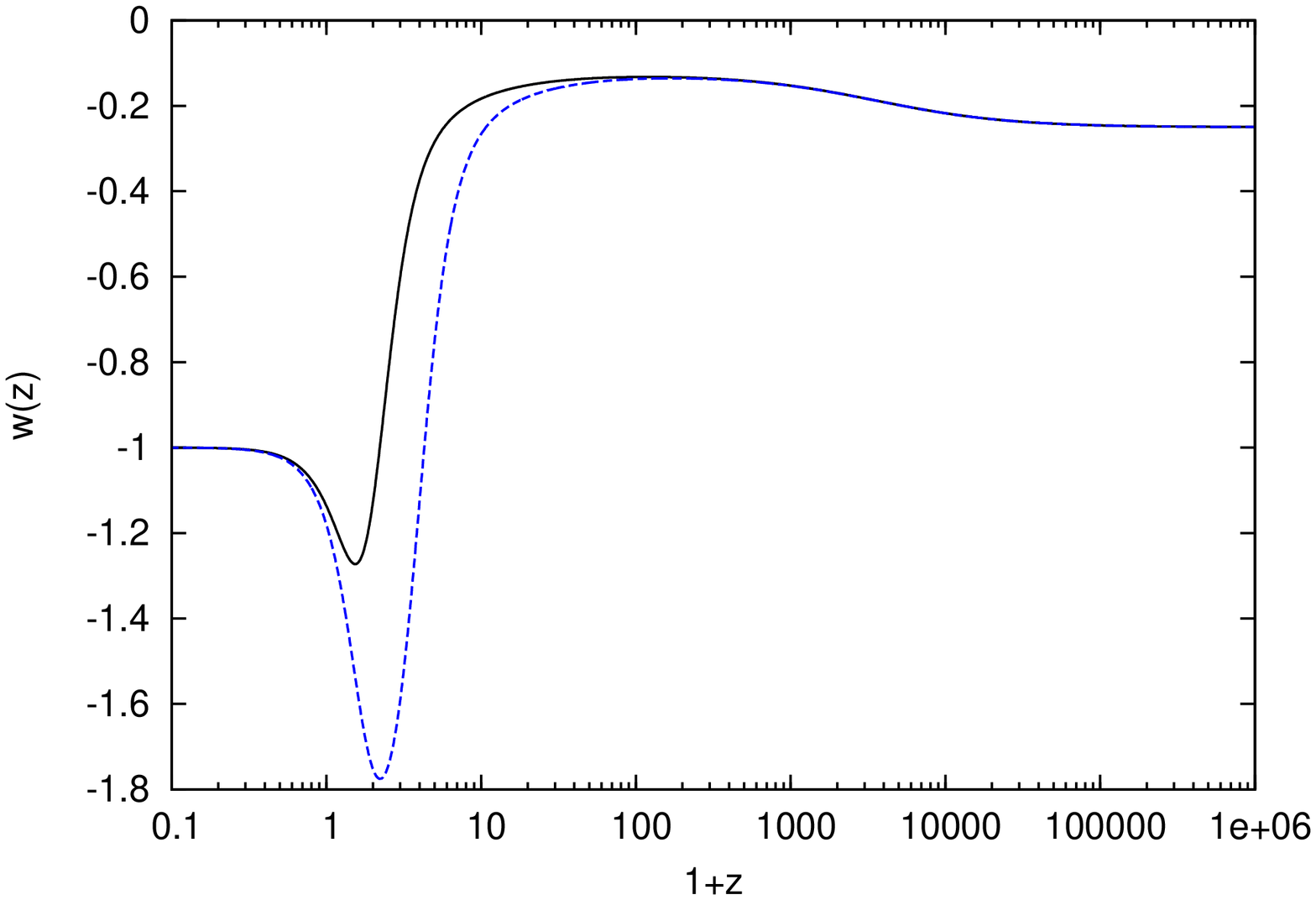}
\caption{Gravitational coupling deviation $\geff/G_N$ and effective 
equation of state $w$ are shown for examples with high redshift initial 
conditions $\rho_{\pi,i}=10^{-6} \rho_{\rm m,i}$ (dashed) and 
$\rho_{\pi,i}=3\times10^{-5} \rho_{\rm m,i}$ (solid).  
Note that adjusting $\rho_{\pi,i}$ to lessen deviations 
in gravity increases the deviations in equation of state, and vice versa. }
\label{fig:tensionw}
\end{figure}

Figure~\ref{fig:tensionw} illustrates this tension between the growth 
history and expansion history behaviors in Galileon cosmology.  The question 
is whether there is a safe path between the monsters.  This requires exact 
numerical computation, scanning over the Galileon parameter space with 
Markov Chain Monte Carlo techniques.

\section{Cosmological Constraints} \label{sec:fit}  

We first emphasize that when using only distance data 
constraints, for example those arising from the cosmic microwave background 
(CMB), baryon acoustic oscillation (BAO) and supernovae, we can 
find an acceptable fit, with a maximum likelihood comparable to 
$\Lambda$CDM (this was also found by \cite{defeliceLCDM}).  Similarly, by 
using only growth 
data one can also find a viable parameter region (cf.\ the low initial 
density curve in Fig.~\ref{fig:tensionw}).  However, these two regions may 
be disjoint and the tension within the combined data constraints forces 
even the best fit to have a poor joint likelihood.  

The second issue is that certain parts of parameter space are restricted 
theoretically due to ghosts or instabilities, as discussed in \cite{applin}.  
Indeed the best fit regions seem to tend to lie close to these because the 
best fits take 
advantage of the near cancellations between terms that can also lead to 
pathologies. We only apply the theoretical criteria to the past behavior 
of the field, that is for $z>0$, since we have no reason to rule out 
models based on their future (observationally untested) behavior.  
Imposing the restrictions at all times, including the future de Sitter 
state, would further constrain the allowed region, increasing the tension 
further. 

For each point in parameter space we solve for the effective dark energy 
equation of state ratio $w(z)$ and the gravitational coupling $\geff(z)$ 
using Eqs.~(18)-(19) and (24) of \cite{applin}.  To stay close to 
quantities best constrained by data, we use $\rhoi$ rather than $x_i$ and 
$\Omega_{de,0}$ rather than $c_2$ as parameters, together with $c_3$, 
$c_4$, $c_5$, and $H_0$.  We adopt a theory or inflationary prior of a 
spatially flat universe.  
We carry 
out the Markov Chain Monte Carlo analysis of the full likelihood surface using 
CosmoMC \cite{cosmomc} as a generic sampler.  The likelihood is given 
by the sum 
\begin{equation} 
{\cal L} = {\cal L}_{\rm CMB} +{\cal L}_{\rm SN}+{\cal L}_{\rm BAO} + 
{\cal L}_{\rm growth} \ . 
\end{equation}

We use the latest observational data to constrain the model.  CMB data from 
WMAP7 is applied in the form of the covariance matrix for the shift 
parameter, acoustic peak multipole, and redshift of decoupling 
\cite{komatsu11}.  Since the Galileon model acts like the standard cosmology 
in the early universe these quantities basically measure the distance 
to last scattering and the matter density.  Distances from 
Type Ia supernovae in the Union2.1 data compilation \cite{union21} 
constrain the expansion history at $z\approx0-1.4$.  Distances from the 
baryon acoustic oscillation feature in the galaxy distribution, measured 
to 6 redshifts at $z=0.1-0.7$ \cite{blake11082635}, probe a somewhat 
different cosmological parameter combination.  For growth constraints 
we use measurements of the growth rate from the WiggleZ survey at four 
redshifts $z=0.2-0.8$ \cite{blake11042948}, and from the BOSS survey at 
$z=0.57$ \cite{reidboss}, in the form of their Eq.~(18) $3\times 3$ 
covariance matrix including expansion quantities, plus the $E_G$ growth 
probe \cite{zhang,reyes}.  
Note that the main conditions needed to apply 
these growth data analyses to constrain a modified gravity model -- 
that the standard $z\gtrsim1000$ matter transfer function and initial 
conditions are preserved, and that growth is scale independent over the 
relevant length scales -- are satisfied by the Galileon case.  
We only use data within this scale independent range, which is
 below the Hubble scale $\sim 4000$Mpc and above the Vainshtein scale $\sim 1$Mpc. 
A thorough analysis of non-linear effects, specifically the Vainshtein screening mechanism, is beyond the scope of this 
paper, although we expect them to be unimportant on scales relevant to linear perturbations.

The results of the MCMC indicate the Galileon model is severely 
disfavored.  
The best fit yields a minimum $\Delta\chi^2=31$ with respect to the best fit 
\lcdm\ model, despite the Galileon case having 4 extra fit parameters. 
We conclude that the entire parameter space of the standard Galileon theory is
strongly disfavored.  The CMB distance to last scattering deviates by 
$\sim 3\sigma$ from the best fit \lcdm\ case, and the individual lower 
redshift distances and growth predictions are similarly in moderate 
disagreement with \lcdm.  The highest impact 
individual constraint arises from the BOSS measurements 
at $z=0.57$.  This leverage bodes well for the impact of future 
redshift surveys on testing gravity on cosmic scales.  The combination 
of all the data leads in aggregate to a poor fit.  

\begin{figure}[htbp!]
\includegraphics[angle=-90,width=\columnwidth]{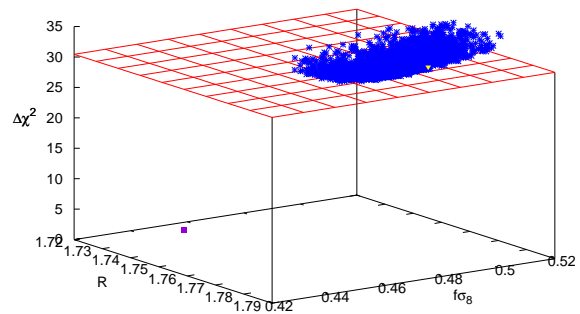}
\caption{$\Delta\chi^2$ relative to the best fit \lcdm\ model (purple square) 
is shown for the Galileon model as a function of the CMB shift parameter 
$R$ (an example of expansion) and growth rate $f\sigma_8(z=0.57)$ (an 
example of growth, measured by BOSS \cite{reidboss}).  
Blue stars are points derived from the MCMC chains, outlining the rough 
paraboloid of the $\chi^2$ surface, with the yellow triangle the best fit.  
No values of the 
Galileon parameters provide a fit with $\Delta\chi^2<30.7$ (horizontal grid). 
}
\label{fig:Rfsig}
\end{figure}

Figure~\ref{fig:Rfsig} shows the $\chi^2$ surface relative to the best 
fit \lcdm\ result in the plane of the CMB shift parameter $R$ and growth 
rate $f\sigma_8(z=0.57)$.  These quantities serve as examples of 
cosmic expansion and growth, respectively.  The $\Delta\chi^2$ is large 
($\chi^2_{\rm gali}=587.2$ to $\chi^2_{\Lambda CDM}=556.5$ for the best fits) 
and the Galileon values are shifted considerably in attempting to fit 
the expansion and growth simultaneously (high yellow triangle for Galileon 
gravity vs low purple square for $\Lambda$CDM).

The key tension between expansion and growth will be more fully 
realized with more accurate data.  If as an example of future data 
we merely change 
the SN data implementation to employ the Union2.1 statistics-only error 
matrix, rather than the full systematics matrix, the improved distance 
measurements exhibit the tension much more clearly, leading to a 
minimum $\Delta\chi^2=53$.  This demonstrates how upcoming supernova 
surveys will also deliver substantial cosmological leverage. 
    
The maximum likelihood values are highly stable with respect to variations 
in the prior ranges.  The parameters $\Omega_{m,0}$, $h$, and 
$\rho_{\pi,i}$ are well constrained (with best fits for the Galileon 
cosmology at 0.302, 0.714, and $\ln(\rho_{\pi,i}/\rho_{m,i})=-11.09$, 
respectively), but the coefficients $c_3$, $c_4$, $c_5$ have strong 
covariance.  
The best fit to observational data minimizes the deviations 
in growth and expansion relative to $\Lambda$CDM, requiring a delicate 
balance among those Galileon coefficients. 
While the nominal best fits are respectively $-2.10$, $-1.71$, $-1.77$, 
there is a long, narrow region of degeneracy.  
(We actually also run extended ranges with logarithmic priors 
to ensure we are not missing a better fit.  Likelihood indeed decreases 
for values of $c_n$ with amplitudes much less than or greater than 1.)  
The degeneracy is moot, however, since the maximum likelihood is so poor.  

\begin{figure}[!thbp]
\includegraphics[angle=-90,width=\columnwidth]{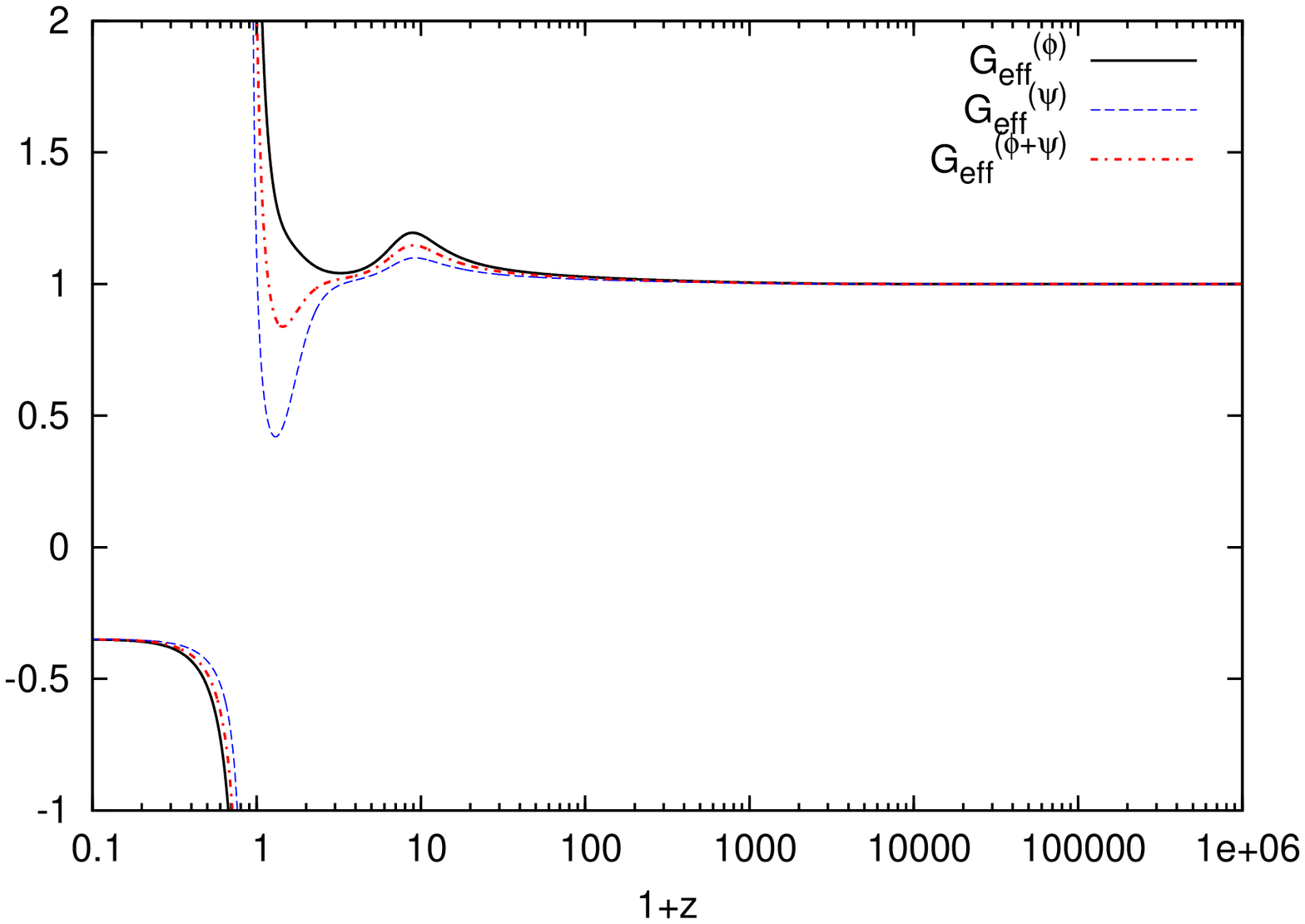}
\includegraphics[angle=-90,width=\columnwidth]{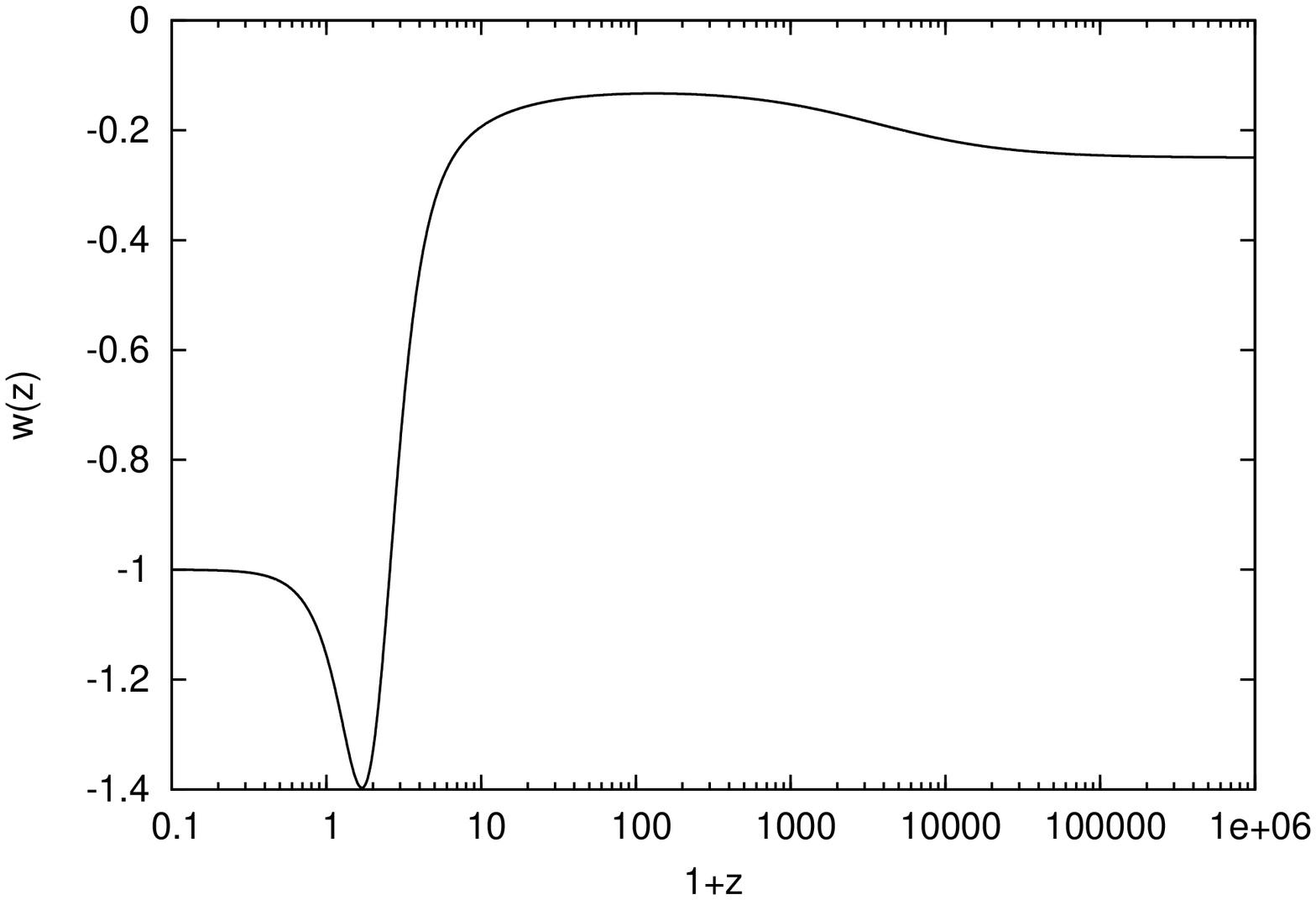}
\caption{Gravitational strength $\geff(z)$ and effective dark energy 
equation of state $w(z)$ are plotted for the Galileon model that best fits 
the current data.  Different $\geff$ superscripts correspond to the 
different modified Poisson equations in \cite{applin}. 
}
\label{fig:best}
\end{figure}

No point in the Galileon parameter space gives a reasonable fit to current 
data.  Moreover, the best fit, poor though it is, is achieved by balancing 
on the edge of a precipice: the gravitational strength diverges in the very 
near future.  To suppress deviations in growth and expansion simultaneously 
the Galileon terms are forced into a highly delicate, and temporary, 
cancellation.  
(Note the divergence of $\geff$ may be ameliorated 
by effects beyond sub-horizon, linear perturbation theory.)  
Since we only 
applied our 
instability criterion to the past, where there is data, we do not rule 
out this model on theoretical grounds despite its Laplace 
instability (negative sound speed squared, $c_s^2<0$, for the dark 
energy perturbations) in the future. 
Figure~\ref{fig:best} exhibits the gravitational strength 
and effective dark energy equation of state as a function of redshift 
for the best fit Galileon model.

\section{Conclusions} \label{sec:concl} 

General relativity has passed all tests to date but lacks a clear 
explanation of the magnitude of the cosmological constant, or origin of 
dark energy, needed to account for cosmic acceleration.  
Two important, pressing questions are whether a sound alternative theory 
of gravity can explain this, and what leverage exists from current 
cosmological data to test such theories. 

Galileon scalar 
fields, which have strong ties to higher dimensional gravity theories, 
can give rise to late time cosmic acceleration 
and possess well behaved, second order field equations with 
symmetries protecting against high energy physics renormalizations. 

We analytically identify, and numerically quantify, a tension, however, 
between Galileon predictions for 
the cosmic expansion history and growth history that severely disfavors 
Galileon cosmology.  Confronting the entire class of standard, uncoupled 
Galileon theory with current observations demonstrates that the predictions 
are a worse fit than general relativity with a cosmological constant by 
$\Delta\chi^2>30$.  If one wanted to abandon the theory prior of spatial 
flatness, \cite{defeliceLCDM} found that adding a free parameter for 
curvature improved the best fit $\chi^2$ by little more than one, and so 
would not have a significant effect on our conclusions. 

In this work we have focussed on the case in which the scalar field 
is not directly coupled to ordinary matter. In a previous paper \cite{applin}
we studied the cosmological evolution in the presence of an
explicit coupling. This previous work highlighted the existence of 
 ghost and Laplace instabilities when a coupling is introduced, however
an exhaustive scan of the parameter space for the coupled models
remains to be undertaken (see \cite{Barreira:2012kk} for work in this direction).

It is striking and signficant that already with current data we can rule out 
an entire, theoretically viable class of extended gravity, one with several 
attractive features. We also established 
that forthcoming data will be able to strengthen these limits to 
$\Delta\chi^2>50$.  More generally, the next generation of cosmological 
measurements will shed strong light on the distinction between modified 
gravity vs general relativity plus a physical dark energy, an exciting 
advance in understanding our universe.

\acknowledgments

This work has been supported by World Class University grant 
R32-2009-000-10130-0 through the National Research Foundation, Ministry 
of Education, Science and Technology of Korea and the Director, 
Office of Science, Office of High Energy Physics, of the U.S.\ Department 
of Energy under Contract No.\ DE-AC02-05CH11231.


\end{document}